# The CMS Silicon Strip Tracker - Overview and Status

**Katja Klein**[*][†]

*1. Physikalisches Institut B, RWTH Aachen, Germany*

*E-mail:* `katja.klein@cern.ch`

With an active silicon area of more than $200\,\mathrm{m}^2$, the silicon strip tracker of the CMS experiment, one of the experiments currently under construction for the future Large Hadron Collider at CERN, will be by far the largest silicon tracker in the world. Efficient mass production and rigid quality control are crucial to finish the detector, comprising of more than 15000 silicon strip modules, in time with optimal quality. After excellent performance of substructures has been proven in various test beam experiments, integration of the active detector elements into the mechanical support structures as well as cabling and testing of these integrated structures has now started.

In this article the CMS silicon strip tracker is described. Production, quality control and integration procedures are outlined and the detector status is reviewed. The detector performance in a recent test beam experiment is summarized.

---

[*]Speaker.

[†]On behalf of the CMS Tracker Collaboration.

## 1. The CMS Silicon Strip Tracker

With its more than 15000 silicon modules and a total active detector area of about $200\,\text{m}^2$, the CMS Silicon Strip Tracker [1, 2] will be the largest silicon tracker ever built. The tracker will have a length of 5.4 m and a diameter of 2.4 m. It will be operated at a temperature of below $-10^\circ$ C.
The active detector elements, the silicon modules, consist of four main parts: a support frame, a Kapton strip which delivers the bias voltage to the sensor back plane and insulates the sensor against the frame, the hybrid with the front-end electronics, and one or two silicon sensors [3]. The sensors are single-sided microstrip detectors with AC-coupled p-type strips in an n-type bulk. Modules mounted with a radial distance of less (more) than 60 cm from the beam line have 320 $\mu$m (500 $\mu$m) thick sensors. More than 20 different module geometries exist, with strip pitches ranging from 80 $\mu$m to 205 $\mu$m. Both single-sided and double-sided modules (two single-sided modules mounted back to back with a stereo angle of 100 mrad) are used.
A cross-section of one quarter of the detector is shown in Fig. 1. The detector is composed of four basic subdetectors. The inner barrel, TIB, consists of four cylindrical layers, with strings of three thin modules mounted inside and outside of the layer half-shells. The inner disks, TID, are composed out of three disks per side, with three rings of modules per disk. The outer barrel, TOB, is constructed out of six cylindrical layers. The basic substructure of the TOB is a rod: a carbon fiber (CF) support frame, which carries either three double-sided or three single-sided thick modules on each side. Finally, each of the two end caps (TECs) consists of nine CF disks. The basic substructure of the TEC is a petal, a wedge shaped CF support plate which carries up to 28 modules arranged in up to seven radial rings. On each disk eight front and eight back petals (which are of slightly different geometry and carry different numbers of modules) are mounted.
The signals of the 512 or 768 silicon strips of a module are processed by four or six APV25 readout chips [4] that are mounted on the front-end hybrid (FE-hybrid) [5]. The APV25 is built in radiation hard 0.25 $\mu$m CMOS technology. It is a 128 channel chip which samples at a frequency of 40 MHz and implements a charge-sensitive amplifier with a time constant of 50 ns, a shaper and a pipeline memory. After arrival of a first level trigger the signals are processed by an analogue circuit, for which two operation modes can be chosen: in peak mode only one data sample is used, while in deconvolution mode three consecutive data samples are reweighted and summed up [6], which leads to a much shorter pulse and thus to correct bunch crossing identification in the high luminosity running phase of the LHC. The signals of two chips are multiplexed onto one data line. The electrical signals are then converted to optical signals in dedicated Analog-Opto Hybrids, and transmitted to the counting room, where they are digitized and processed by 10 bit ADCs.

## 2. The Status of the CMS Silicon Strip Tracker

In this chapter, the status of the tracker as of July 2005 is described.
In the past, several problems had to be faced with two important module components, the silicon sensors and the FE-hybrids. The 24244 sensors are delivered by two companies, ST Microelectronics (STM) (Catania, Italy) and Hamamatsu Photonics K.K. (HPK) (Hamamatsu-City, Japan), to five quality control centers. There 5% of all sensors plus all sensors from suspicious batches are characterized in detail, with all important sensor parameters such as the interstrip capacitance and depletion voltage being measured. Furthermore 1% of the sensors are subjected to irradiation tests with neutrons and protons to monitor the sensor processing quality. In addition the longterm behaviour of several per cent of sensors is studied. This elaborate quality control reveiled several flaws (unstable leakage current, oxidation) of the sensors of STM. In reaction to this the bulk of the production was shifted to HPK to assure the high quality of the modules. As of July 2005, all thin and about 85% of the thick sensors are delivered and qualified.

For the FE-hybrids, two companies are involved: the flex circuits are produced by Cicorel SA (Boudry, Switzerland), the components are assembled at Hybrid SA (Chez-le-Bart, Switzerland). The hybrids are tested directly at the company with a fast test, and later at CERN prior to the bonding. Several severe problems have been spotted, e.g. in certain batches thin vias developed a bad electrical contact. This problem was solved by an improved production process. About 400 good hybrids are delivered per week, and 70% of all hybrids have been delivered and accepted.

The modules themselves are glued with a precision of a few tens of microns on six fully automatic pick-and-place robots. Up to 20 modules can be glued per robot per day, and 99% of the modules built are within specification. The modules are then wire bonded using more than 20 automatic bonding machines with a throughput of more then five modules per day and machine. The module quality is excellent, with a few per mille of bad strips. About 35% of modules are built.

The modules are then integrated onto the petal and rod structures (or, in the case of the TIB, directly onto half-shells). For the TEC, up to 28 Analog-Opto Hybrids and modules have to be mounted onto the bare petals. The assembled petals are then characterized in a longterm test during six thermo-cycles between CMS operating temperature and room temperature. Pedestals are taken at both temperatures and the petals are graded basically according to the number of bad strips and the leakage current. The first experience with 5% of petals already built is encouraging: only a few per mille of bad strips are present per petal.

Fully equipped petals and rods are then shipped to the integration centers, where they are integrated into the mechanical support structures. Both the TOB and TEC structures are assembled with a precision of the order of $200\,\mu$m. The TEC structures are fully equipped with optical ribbons. Successful trial integrations of both petals and rods have been performed. The final integration of petals and rods into the structures is foreseen to start end of 2005. For the TIB three out of eight CF cylinders are mechanically ready (precision of the order of $200\,\mu$m), and the first single-sided layer has been fully integrated with modules. The readout test and burn in at low temperature is ongoing. First measurements indicate less than 1% of bad strips, and a noise level better than measured in previous test beams.

## 3. Test Beam Performance of Tracker Substructures

In June 2004 TOB and TEC substructures have been operated in pion ($120\,\text{GeV}/c$) and muon particle beams at CERN. The TEC setup consisted of a front and a back petal (51 modules). Corresponding to about 1% of the TEC system, this was the largest substructure of the CMS tracker tested so far. Stable system operation was achieved both at room temperature and at CMS operating temperature. The signal-over-noise (S/N) distributions of all modules have been measured in peak and deconvolution mode both in the cold environment and at room temperature, and the mean value has been calculated per module geometry. In the cold, most probable values of the S/N of 28-33 (35-42) and 19-22 (20-24) have been observed for thin (thick) sensors in peak and deconvolution mode, respectively. The Equivalent Noise Charge (ENC) was studied as a function of the capacitance (Fig. 1), which depends for CMS sensors only on the strip length [2]. The ENC is found to be 714-1155 (1020-1681) electrons in peak (deconvolution) mode, in good agreement with the expectation based on measurements of the noise of the APV25 chip. A mean common mode noise of $(173\pm38)$ electrons in peak and $(299\pm76)$ electrons in deconvolution mode was measured. As expected, a decrease of the noise of about 10% was observed between room temperature and CMS operating conditions.

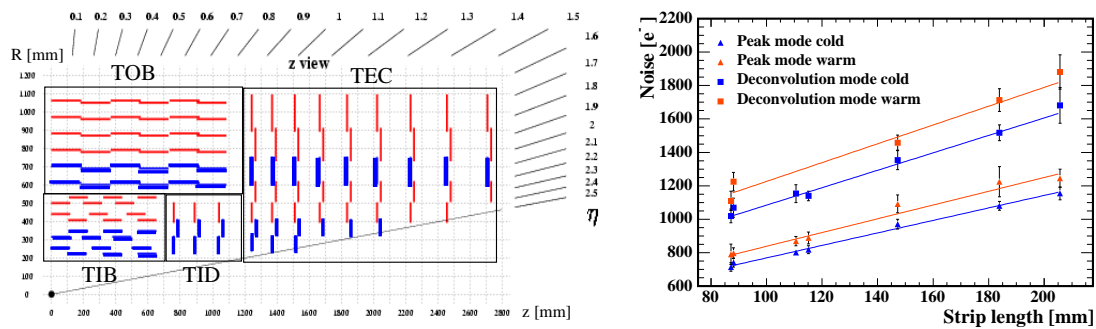

**Figure 1:** Left: cross-section of one quarter of the tracker in the longitudinal view. Thick (thin) lines represent double-sided (single-sided) modules. Right: ENC versus strip length for peak and deconvolution mode and the two operating temperatures. Each entry represents the mean value of all APVs of a certain geometry. The uncertainties correspond to the RMS of the resulting distributions.

## 4. Summary

After the technical problems with sensors and hybrids have been overcome, and with the excellent performance of tracker substructures proven in test beam experiments, the module production is now running smoothly at full swing and is targeted to finish in early 2006. Tracker integration has started, and the tracker is scheduled to be integrated into the CMS detector in autumn 2006.

## References

[1] CMS Collaboration, *The Tracker Project, Technical Design Report*, CERN/LHCC 98-6, CMS TDR 5 (15. April 1998).

[2] CMS Collaboration, *Addendum to the CMS Tracker TDR*, CERN/LHCC 2000-016, CMS TDR 5 Addendum 1 (21. April 2000).

[3] L. Borello, E. Focardi, A. Macchiolo and A. Messineo, CMS Note 2003/020 (2003).

[4] G. Cervelli *et al.*, Nucl. Instrum. and Methods **A466** (2001) 359.

[5] U. Goerlach, *Industrial production of Front-End Hybrids for the CMS Silicon Tracker*, Proc. of the 9th Workshop on Electronics for LHC experiments, Amsterdam, Netherlands (2003).

[6] S. Gadomski *et al.*, Nucl. Instrum. and Methods **A320** (1992) 217.